\documentstyle[11pt,newpasp,twoside,epsf]{article}
\def\edcomment#1{\iffalse\marginpar{\raggedright\sl#1\/}\else\relax\fi}
\reversemarginpar

\begin{document}
\title{The Impact of Massive Stars on the Formation of Young Stellar Clusters}
\author{S. T. Megeath, B. Biller, T. M. Dame, E. Leass, \& R. S. Whitaker }
\affil{Harvard--Smithsonian Center for Astrophysics, 60 Garden St, Cambridge,
 MA 02138, USA}
\author{T. L. Wilson}
\affil{SMTO, Steward Observatory, University of Arizona, Tucson, AZ 85721, USA \& MPIfR, Postfach 2024, D-53010, Bonn, Germany}

\begin{abstract}

Massive OB stars play an important role in the evolution of molecular
clouds and star forming regions.  The OB stars both photo--ionize
molecular gas as well as sweep up and compress interstellar gas
through winds, ionization fronts, and supernovae.  In this
contribution, we examine the effect massive stars have on the
formation of young stellar clusters.  We first discuss the processes
by which OB stars destroy cluster--forming molecular cores, and hence
terminate star formation.  We overview observational evidence that OB
stars forming in young stellar clusters destroy their parental cores
on a timescale of 0.1 Myr, and we discuss some of the implications of
this result.  We then summarize extensive observations of the NGC 281
molecular cloud complex, and present evidence that two types of
triggered star formation are occurring in this complex.  Our goal is to
underscore the impact massive stars have on cluster formation over
distances ranging from 0.1 pc to 300 pc.

\end{abstract}

\section{Introduction}

Stars are formed not in isolation, but in associations, groups and
clusters. Associations span tens of parsecs and are defined by the
molecular cloud complexes from which their constituent stars form;
some examples are the T association in the Taurus molecular cloud, and
the extensive association of OB stars interacting with the Orion
molecular clouds.  Within these molecular clouds, star formation is
typically not widespread, but concentrated into distinct,
parsec--sized groups and clusters with memberships ranging from ten
stars to a thousand stars\footnote {Since systems of more than 100
young stars may persist as gravitationally bound open clusters, Adam
\& Myers (2001) define clusters as systems of more than 100 stars, and
groups as systems containing less than 100 stars.}.  The stars in
associations and clusters are not formed simultaneously in a single
star forming event, but over a span of 1 Myr in clusters (Palla \&
Stahler 2000), and a span of several to 10 Myr in associations
(Brice\~no et al. 2001).  The immediate implication of both the
non-isolated and non-coeval nature of star formation is that the
formation of stars can be affected by winds, radiation, and outflows
from previous generations of stars.

OB stars have the most dramatic influence on star formation. These
stars both ionize and photodissociate molecular clouds with their
intense UV radiation and sweep up atomic and molecular gas with winds,
ionization fronts, and -- at the end of their lives -- supernovae.  In
particular, stars with spectral types of B2 or earlier produce enough
UV radiation to effectively ionize cluster--forming molecular
cores. Since OB stars are relatively rare, it is important to assess
their importance by examining unbiased near--IR surveys of molecular
clouds.  In an analysis of the 2MASS 2nd incremental release data
covering the Perseus, MonR2, Orion A and Orion B clouds, Carpenter
(2000) found a total of 1200 isolated stars, 200 stars in eight
groups, and 2900 stars in six clusters.  Hence, the majority of the
stars in these clouds formed in clusters containing 100 or more stars.
In the sample, four of the clusters, GGD12-15, Mon R2, NGC 2024 \& the
Orion Nebula Cluster, contain at least one star with a spectral type
of B2 or earlier.  One of the groups, NGC 2068\footnote{Our
classification of NGC 2068 as a group is due to the partial coverage
of this region in the 2nd incremental release of the 2MASS data.
While Carpenter finds only 45 stars in the incomplete 2MASS data, Lada
(1989) reports 192 stars in the region, indicating that NGC 2068
should be classified as a cluster.}, contains a B1.5 star.  In total,
57\% of the stars detected in the studied regions are in close
proximity to ionizing OB stars.  This number is a lower limit since it
only counts OB stars which form within clusters. As we will
demonstrate, external OB stars may also influence the formation of
clusters.

In this paper, we discuss how massive stars influence the process of
cluster formation on sizescales ranging from 0.1 pc to 300 pc, with a
particular emphasis on our own study of the NGC 281 region.  We start
by examining the disruption of cluster--forming molecular cores by
nascent OB stars within the cluster.  As we demonstrated above, the
majority of observed young clusters appear to contain OB stars.  We
then study the disruption of cluster forming cores by neighboring OB
stars, which are situated a few parsecs away from the star forming
region.  This case occurs during the formation of an OB association, when
OB stars which have dispersed their own parental cores begin to ionize
neighboring star forming cores.  Finally, we discuss a possible case
of supernova driven star formation, where the influence of OB stars
extends beyond the boundaries of molecular cloud complexes to
distances of a few hundred parsecs.

\section{Core Destruction by Internal OB Stars ($< 1$~pc)}

The birth sites of massive stars are 1000~M$_{\odot}$ dense molecular
cores with densities of $10^5$~cm$^{-3}$ and sizes of 1 pc (Plume et
al. 1997).  Within these cores, entire clusters of lower--mass stars
form.  OB stars still embedded within their parental cores are thought
to form ultracompact HII regions (UCHIIs), HII regions with sizes of
$< 0.1$~pc and electron densities $> 10^4$~cm$^{-3}$.  This picture is
supported by the growing number of observations showing UCHIIs
surrounded by embedded clusters of lower--mass stars (Megeath et
al. 1996; Tapia, Persi \& Roth 1996; Pratap, Megeath \& Bergin 1999).
After OB stars dissipate their parental dense cores, the HII regions
expand and are no longer classified as UCHIIs (Tieftrunk et
al. 1997).  Hence, the lifetime of UCHIIs is the time required for
young OB stars, once they have formed, to destroy their parental cores
and terminate the process of cluster formation.

The best current estimates of the lifetimes of UCHIIs are based on
empirical arguments.  Wood \& Churchwell (1989) found that UCHIIs can
be identified on the basis of their IRAS colors.  By comparing the
number of IRAS identified UCHIIs in the Galaxy with the observed
number of OB stars in the solar neighborhood, they estimated a
lifetime of 0.5 Myr.  By eliminating lower--mass stars contaminating
the IRAS sample, as well as taking into account variations in the
density of UCHIIs with galactic radius, this lifetime has been revised
by Comeron \& Torra (1996) and Casassus et al. (2000) to $0.05 - 0.1$
Myr.

\begin{figure}
\plotfiddle{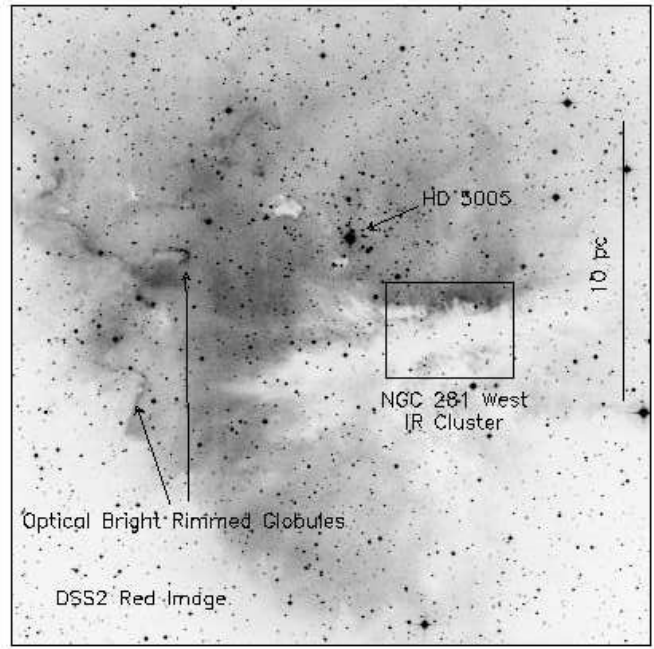}{2.5 in}{-90}{60}{60}{-190}{235}
\vskip 0.75 in
\caption{2nd Generation Digital Sky Survey, red band image of the NGC
281 Nebula and IC 1590 Cluster. The primary ionizing star is HD 5005,
an O5.5 star.  The eastern side of the nebula shows numerous bright
rimmed globules, but the most active region of ongoing star formation
is the obscured IR cluster towards the southwestern quadrant of the
nebula.}
\end{figure}

The estimated UCHII lifetimes indicate that once an OB star is formed
in a cluster forming core, the core destruction occurs on a timescale
of 0.1~Myr.  Since the timescale for cluster formation is ten times
longer, 1 Myr (Palla \& Stahler 2000), only $1/10$ of the forming
clusters should contain ultracompact HII regions.  Interestingly, in
the combined surveys of W3 (Tieftrunk et al. 1998), Perseus, Orion A,
Mon R2 (Carpenter 2000) \& Orion B (Lada 1992), $4/10$ of the clusters
contain UCHIIs: Mon R2, W3 OH, W3 Main, and GGD12-15.  Although the
ratio is four times that expected, the current sample of molecular
clouds is small.  A larger sample is needed to firmly establish the
fraction of star--forming clusters with UCHIIs.

\section{Core Destruction by External OB Stars (1-10 pc)}

\begin{figure}
\plotfiddle{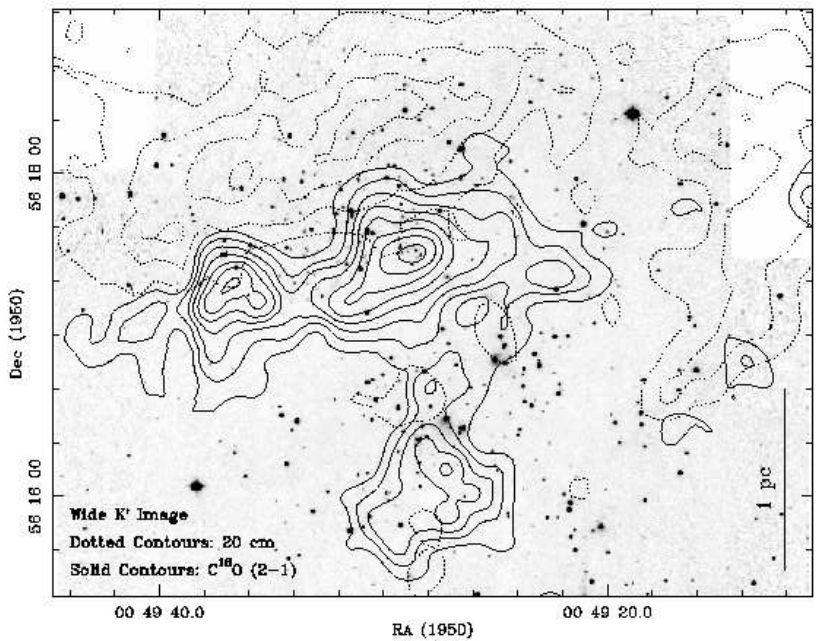}{2.5 in}{-90}{55}{55}{-200}{230}
\vskip 0.75 in
\caption{Greyscale: 2~$\mu$m image of the NGC 281 West cluster. Dotted
contours: 20 cm VLA data tracing the ionized gas. The contour levels
are 1.8 to 5.4 by 1.8 mJy/beam. Solid contours: C$^{18}$O ($2
\rightarrow 1$) IRAM 30~m map of the molecular gas in the interface.
The contours levels are 5.24 to 15.72 by 1.747~K~km~s$^{-1}$. The
ionizing star, HD5005, is 4 pc to the northeast of the molecular cores
(outside the near--IR image).  Note that the cluster is divided into 
northern and southern sub-cluster.  }
\end{figure}

Optical images of HII regions commonly show bright rimmed and cometary
globules: dense molecular gas that survives in the HII region once the
surrounding, lower density gas is swept away.  These globules are
directly exposed to the UV radiation from OB stars within the HII
region.  In a process called photoevaporation, the UV radiation
produces a flow of ionized gas off the surfaces of the clump; the
optical line emission producing the bright rims arises in this ionized
flow.  Despite the continual erosion of the molecular gas, a
theoretical analysis by Bertoldi \& McKee (1990) has demonstrated that
these globules can survive for millions of years in the HII region,
depending on their mass and distance from the OB stars.  This
relatively long timescale for the core destruction allows for the
formation of groups and clusters in the exposed cores.  Several
examples of young stellar groups forming in bright rimmed globules
have been reported by Sugitani, Tamura \& Ogura (1995).

The formation of groups and clusters in externally ionized cores may
not result solely from the long destruction timescales, but also from
the triggering of star formation by the ionization.  A likely example
of triggered cluster formation is the NGC 281 infrared cluster, the
richest cluster yet observed towards a bright rimmed globule.  It is
the most active region of ongoing star formation in NGC 281,
containing at least one massive star, as evidenced by a $10^4$
L$_{\odot}$ IRAS source (Henning 1994), and a cluster of 240
lower--mass stars, as detected in near--IR images (Megeath \& Wilson
1997).  The infrared cluster, located in the southwestern quadrant of
the NGC 281 nebula, is not associated with the optical bright rimmed
globules apparent in the nebula (Fig.~1).

\begin{figure}
\plotfiddle{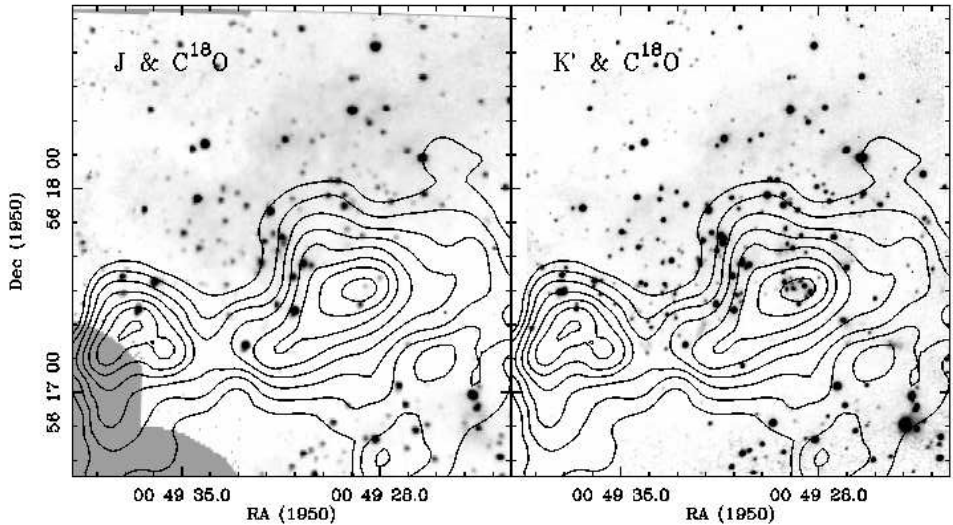}{2.5 in}{-90}{50}{50}{-190}{230}
\caption{Left: $J$-band image of the NGC 281 HII/H$_2$ interface with
contours of the C$^{18}$O ($2 \rightarrow 1$) emission overlaid.
Right: $K$-band image with C$^{18}$O contours.  The images are
centered on the northern sub-cluster, which is associated with the two
northernmost molecular cores. The stars apparent in the bottom right
corner of the images are in the southern sub-cluster associated with the
southernmost molecular core (see Fig.~2).}
\end{figure}

The evidence for triggered star formation comes from combined
centimeter, millimeter, and near--IR data.  Since optical observations
of this region are complicated by a spatially varying extinction, we
obtained C$^{18}$O maps, tracing the dense cores, and a VLA 20 cm map,
tracing the ionized gas, to disentangle the structure of the H$_2$/HII
interface with angular resolutions $< 20''$.  The maps show that the
molecular gas is concentrated into three main cores and several
smaller clumps (Fig.~2).  To the northern edges of each core, 20 cm
emission tracing the ionized gas is detected, with particularly strong
20 cm emission adjacent to the two northernmost cores.  These cores
are {\it radio} bright rimmed globules.  In support of the radio data,
recent Hubble space telescopes images taken with the WFPC2 camera show
the optical emission from H$\alpha$ and [SII] lines extending to the
cores.  These data demonstrate that the cores are exposed to the UV
radiation from the ionizing O star, and are photoevaporating.  Megeath
\& Wilson (1997) estimate that at the current rate of
photoevaporation, the observed cores will be fully evaporated in 2.5
Myr.

A comparison of the pressure from the ionized, photoevaporating gas at
the clump surfaces to the turbulent pressure of the molecular gas in
the clumps, indicates that the pressure of the ionized gas is
sufficient to drive shock waves into the molecular cores.  The shocks
would have a velocity of 1.5~km~s$^{-1}$ and a crossing time of 0.7
Myr. The C$^{18}$O data show a ridge of blueshifted gas along the
northern edge of the cores: this may be the detection of gas swept up
by the shock (see Megeath \& Wilson 1997).

\begin{figure}
\plotfiddle{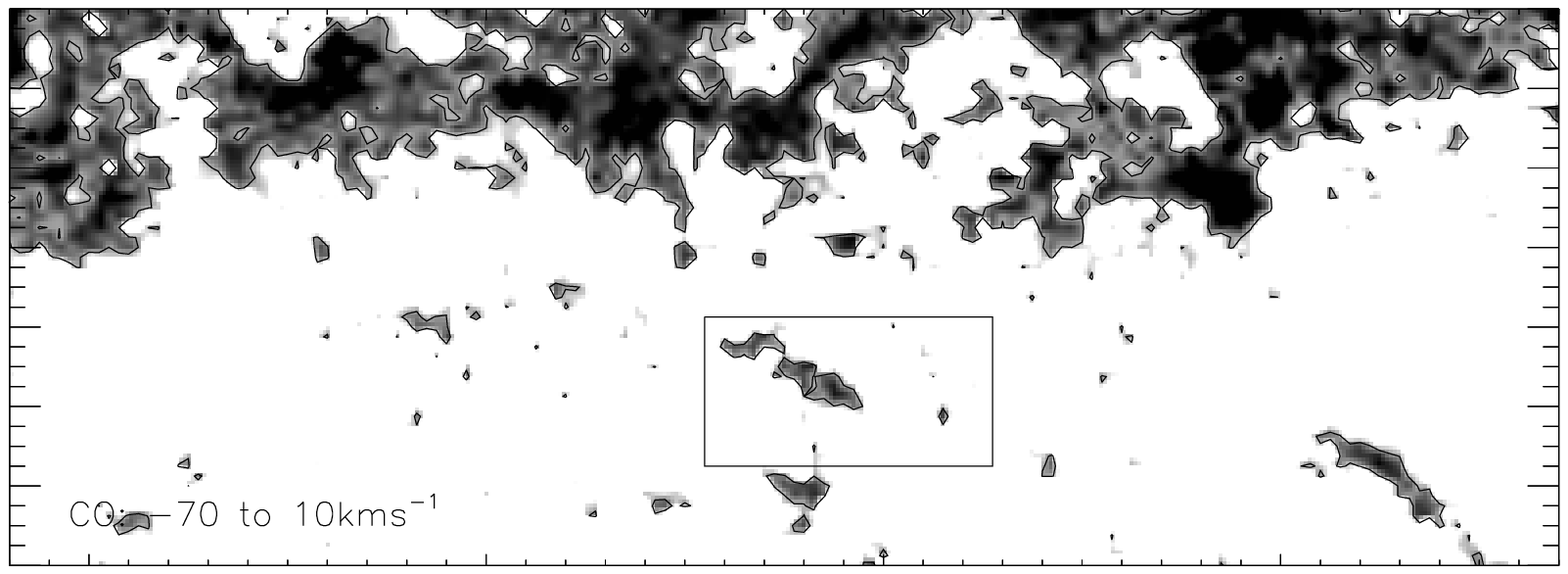}{2.3 in}{0}{55}{55}{-170} {40}
\plotfiddle{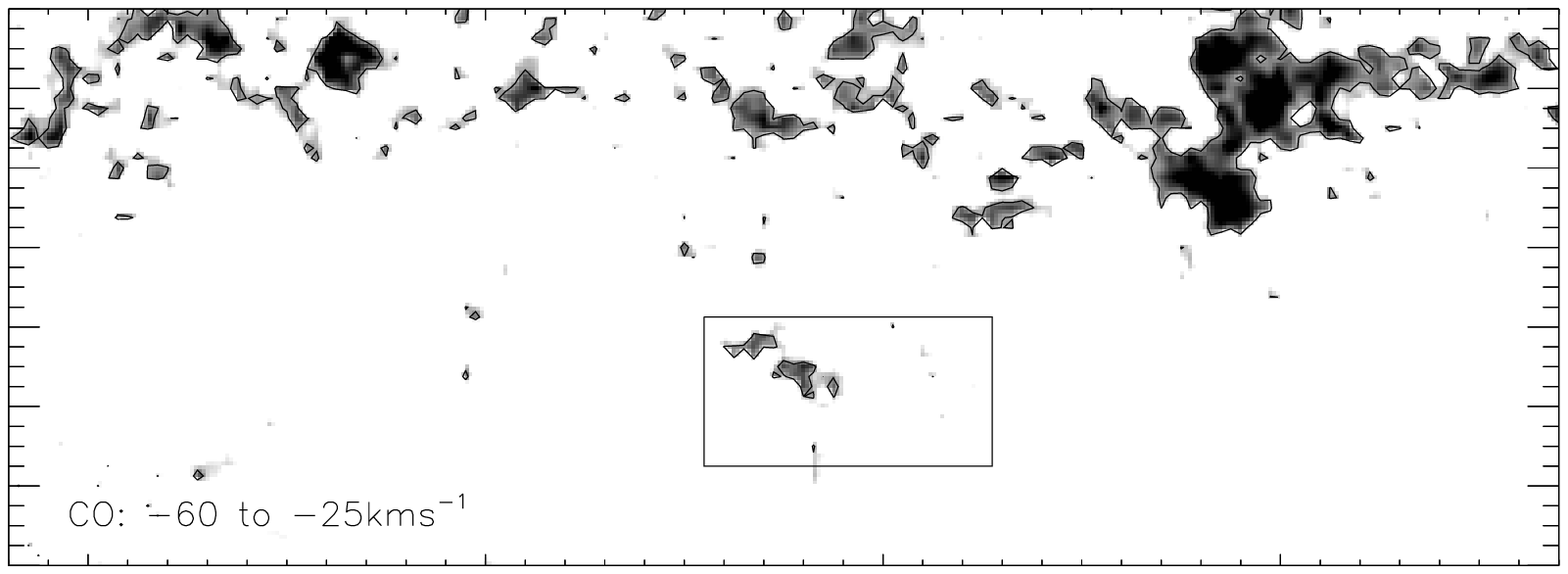}{2.3 in}{0}{55}{55}{-170}{130}
\plotfiddle{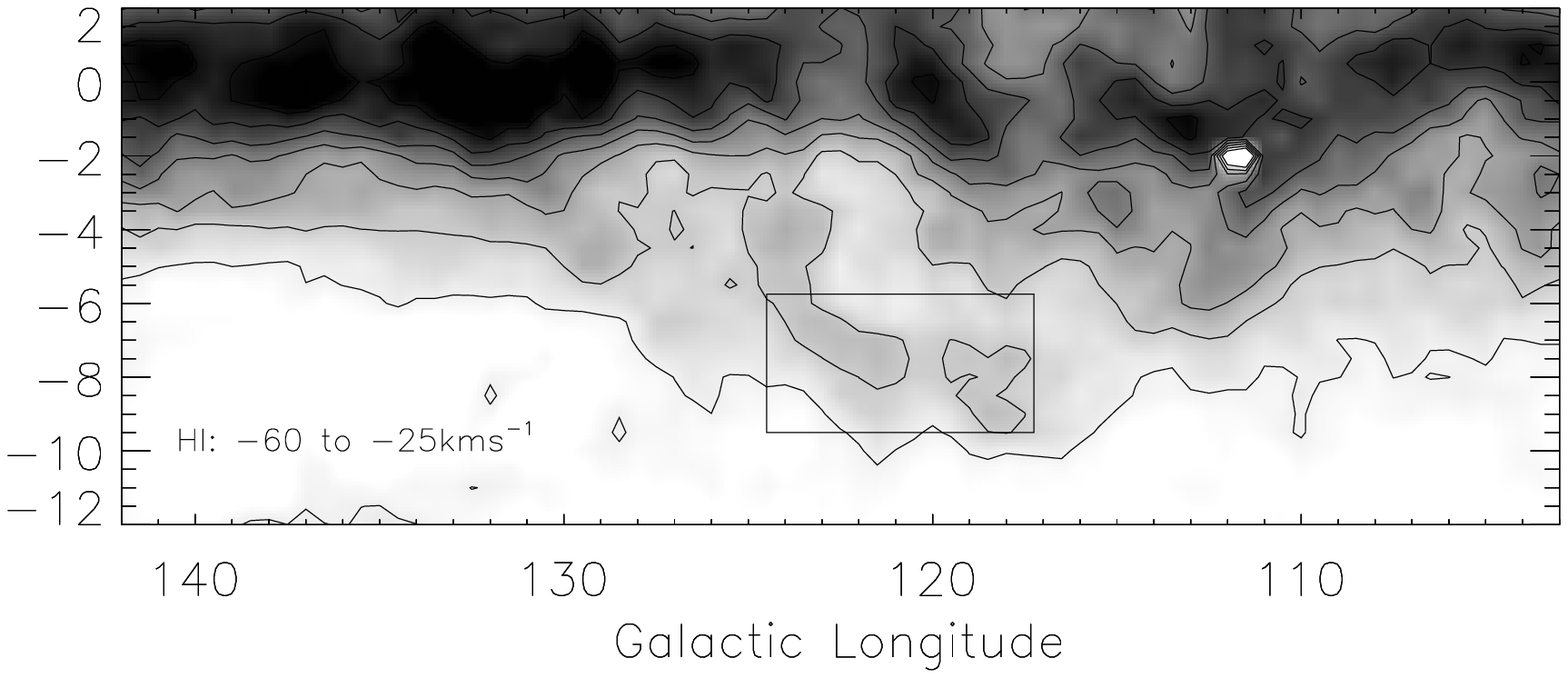}{2.3 in}{0}{55}{55}{-170}{220}
\vskip -3.6 in
\caption{The velocity--integrated CO ($1 \rightarrow 0$) line emission
and HI 21 cm line emission from the CfA CO survey and the
Leiden--Dwingeloo HI survey (Dame, Hartmann \& Thaddeus 2001; Hartmann
\& Burton 1997).  Top: the local and Perseus arm velocity gas.
Middle: The Perseus arm velocity gas.  Bottom: 21 cm map of the HI
emission integrated over the Perseus arm velocities. The NGC 281 cloud
complex is delimited by the rectangle.}
\end{figure}

Near--IR images of the two northermost cores show a steep gradient in
the surface density of stars with most of the embedded stars
concentrated near the photoevaporating surfaces of the cores (Fig.~3).
The lack of a corresponding gradient in the extinction through the
cores, as demonstrated by the C$^{18}$O maps, implies that the
observed gradient in stellar density is due to an asymmetric
distribution of stars.  Megeath \& Wilson (1997) have proposed that
the asymmetry is the result of triggered star formation.  In this
picture, the shock fronts have traversed roughly halfway through the
cores.  As the shocks compress the gas, they precipitate the formation
of stars which then fill the post-shock gas.  Near--IR imaging of
groups in optical bright rimmed globules have detected similar
asymmetries in the distribution of stars (Sugitani, Tamura \& Ogura
1995). These data suggest that shock--triggered formation may be a
common process in cores exposed to the UV radiation from neighboring
OB stars.

\section{Supernovae Driven Formation of Clusters ($> 100$~pc)}

The NGC 281 nebula is part of a complex of molecular clouds in the
Perseus arm of the galaxy (Fig.~4).  The measured distance to the
nebula (2.9 kpc; Guetter \& Turner 1997) implies that the complex is
located 300 pc below the galactic plane; in contrast, the rest of the
Perseus arm clouds are found within 150 pc of the plane.  Another
unusual property of the NGC 281 complex is the high dispersion in
cloud velocities: the longitude-velocity diagram in Fig.~5 shows a
dispersion of 45~km~s$^{-1}$.  The high velocity dispersion suggests
that the complex may only be a chance alignment of clouds in the sky.
However, the lack of other Perseus arm clouds at these galactic
latitudes, the linking of the CO clouds by the HI emission (Fig.~4 and
5), and the elliptical distribution of CO and HI emission in the
longitude-velocity diagram, all support the notion that these clouds
form a bona fide complex of molecular and atomic clouds.

The elliptical pattern evident in the longitude-velocity diagram is
indicative of an expanding ring.  The diameter of the ring is 270 pc,
the expansion velocity in the line of sight is 22~km~s$^{-1}$, and the
resulting dynamical time is 6~Myr.  The mass in molecular gas, using a
standard conversion from the velocity--integrated CO temperatures to
$H_2$ column density, is $10^5$~M$_{\odot}$.  The HI mass integrated
over latitudes of -6$^{o}$ to -11$^{o}$ and velocities of
-15~km~s$^{-1}$ to -60~km~s$^{-1}$ is $3.5 \times 10^5$~M$_{\odot}$.
The resulting kinetic energy is $4.5 \times 10^{51}$~ergs, comparable
to the energy of a supernova.



We are currently exploring the possibility that the ring has formed in
gas lifted out of the plane by the expansion of a superbubble.  The HI
emission in the NGC 281 complex forms part of a loop of HI extending
out of the plane; this loop may trace the shell of a superbubble
driven by supernovae and winds from OB associations near the plane of
the Galaxy (Fig.~4). Models of superbubbles expanding into a
stratified galactic ``atmosphere'' predict that at a distance of a few
hundred parsecs above the plane, the upward expansion of the bubble
accelerates resulting in a blowout (MacLow \& McCray 1988).  Numerical
simulations of blowouts predict that the fragmenting shell can have
velocities as large as 50~km~s$^{-1}$ {\it tangential} to the galactic
plane (MacLow, McCray \& Norman 1989). It is then plausible that the
observed expanding ring has been formed in a blowout, with the clouds
located at the distance from the plane where the blowout started.
This interpretation suggests that the formation of the entire NGC 281
complex has been triggered by previous generations of OB stars.
This mode would operate on a totally different timescale (6 Myr vs 1
Myr) and sizescale (300 pc vs 4 pc) than the triggered star formation
discussed in the previous section.

\begin{figure}
\plotfiddle{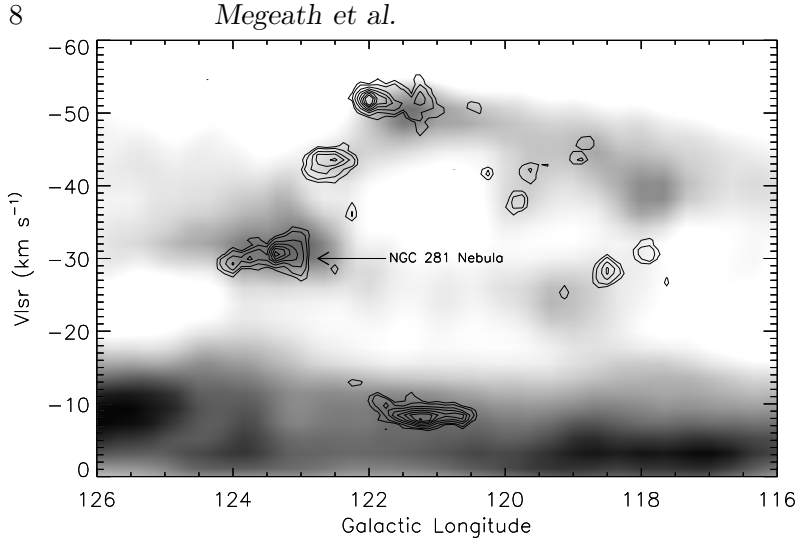}{1.8 in}{0}{50}{50}{-170}{-20}
\vskip 0.4 in
\caption{Longitude-velocity diagram of the CO ($1 \rightarrow 0$)
(contours) and HI 21 cm emission (greyscale) integrated over $b = -8^{o}$ to
$-5^{o}$.  The contours levels are 0.1, 0.2 to 1.4
by 0.2~K~deg.  }

\end{figure}

\end{document}